\documentclass[twocolumn,letterpaper,prl,showpacs]{revtex4}
\usepackage{graphicx}% eps files
\usepackage{bm}% bold math
\usepackage{amsmath}% equation formatting
\usepackage{amssymb}% math symbols
\newlength{\pwd} 
\newlength{\twocolumnwidth}

\begin{document} 
%******************************************* 
\title{Regularisation as a quantised low-pass filter} 
%*******************************************
\author{L.\ I.\ Plimak} 
\affiliation{Abteilung Quantenphysik, Universit\"at Ulm, 
D-89069 Ulm, Germany.} 
%*******************************************
%\author{S.\ Stenholm} 
%\affiliation{\ULM.} 
%\affiliation{\StigAffilSw.} 
%\affiliation{\StigAffilFin.} 
%******************************************* 
\date{\today} 
%*******************************************
\begin{abstract} 
%*******************************************
A divergence-free approach to relativistic quantum electrodynamics based
on regularisation of equations of quantum mechanics is
discussed. This approach is shown to
be exactly equivalent to the conventional Feynman-Dyson renormalisation
techniques. 
%******************************************* 
\end{abstract}
%******************************************* 
\pacs{}
%*******************************************
\maketitle 
%******************************************* 
{\em Introduction.---\/}%
It is appropriate to start this letter with a word of caution. {\em This is not a solution to the problem of infinities in quantum field theory.\/} My goal is to construct a computational scheme with the standard perturbation approach leading {\em directly\/} to regularised diagram series, so that infinities never occur (for all standard definitions see Refs.\ \cite{Schw,PerKeld}). This scheme may be regarded as a constructive implementation of Bogoliubov's point that equations of quantum field theory are mathematically ``underdefined'' and hence contain some freedom. This freedom is implemented by defining ``observable'' fields as mixtures of the physical and fictitious ones. The interaction is then introduced in such a way that divergences never occur. This yields a perturbation expansion which coincides {\em exactly\/} with the Pauli-Villars-regularised diagram series in the conventional approach \cite{Schw}, with the fictitious masses becoming regularisation masses. The standard renormalisation procedure then ``traps'' the freedom introduced into the equations of motion in observable mass and charge of the electron. Unlike in the conventional Feynman-Dyson approach, all expressions remain mathematically meaningful at all stages of calculations. 

%*******************************************
\begin{figure}[t] 
\begin{center} 
\includegraphics[width=0.9\columnwidth]{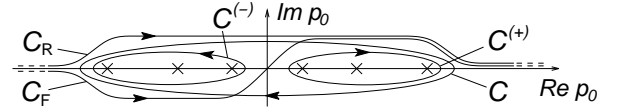}
\end{center} 
\caption{Integration contours in the complex plane of energy. The closed contours encircle all poles of integrands (shown schematically as crosses). Poles on the negative real axis appear in relativistic models. For details see e.g.\ Schweber's textbook \cite{Schw}.\label{CCon}} 
\end{figure} 

%*******************************************
{\em Response and regularisation.---\/}%
%*******************************************
In detail, this connection will be discussed elsewhere \cite{APIII}, see also \cite{AP}. The key idea may in fact be demonstrated for the oscillator with a variable frequency, with the Hamiltonian, 
%=============================================
\protect{\begin{align}{{\begin{aligned} 
\hat H_v(t) = \protect\protect\ensuremath{\big[
\Omega_0 - v(t)
\big]} \hat a_0^{\dag} \hat a_0 . 
\end{aligned}}}%
\label{eq:11a} % \nonumber % &
\end{align}}%
%+++++++++++++++++++++++++++++++++++++++++++++
Here, $\hat a_0,\hat a_0^{\dag}$ is the standard creation/annihilation pair in the Scr\"odinger picture (SP), $[\hat a_0,\hat a_0^{\dag}]=1$, $\Omega_0$ is the oscillator frequency, and $v(t)$ is a c-number function. 
We use units where $ \hbar =c=1$. 
Operators in the interaction picture (IP) will be signalled by the presence of the time argument: $\hat a_0^{\dag}(t),\hat a_0(t)$, and those in the Heisenberg picture (HP) by uppercase letters: $\hat A_0^{\dag}(t),\hat A_0(t)$. These conventions also apply to the field operators below. 

Feynman's diagram technique \cite{Schw} for this toy system comprises two graphical elements, 
%=============================================
\protect{\begin{align}{{\begin{aligned} 
\raisebox{-0.6ex}{\includegraphics{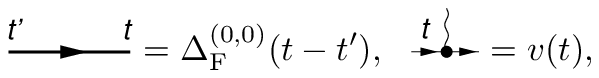}} 
\end{aligned}}}%
\label{eq:16a} % \nonumber % &
\end{align}}%
%+++++++++++++++++++++++++++++++++++++++++++++
where (cf.\ Fig.\ \ref{CCon})
%=============================================
\protect{\begin{align}{{\begin{aligned} 
\Delta _{\text{F}}^{(0,0)}(t) = i\theta(t) \text{e}^{-i\Omega_0 t} 
= \int _{C_{\text{F}}} \frac{d \omega }{2\pi}\, 
\frac{\text{e}^{-i\omega t}}{\Omega_0-\omega } . 
\end{aligned}}}%
\label{eq:1b} % \nonumber % &
\end{align}}%
%+++++++++++++++++++++++++++++++++++++++++++++
The only divergent diagram \cite{endn1} in this technique is the contribution to the vacuum phase, with $\bar v = \int dt v(t)$, 
%=============================================
\protect{\begin{align}{{\begin{aligned} 
\raisebox{-0.6ex}{\includegraphics{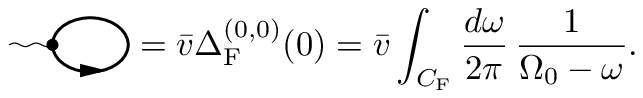}} 
\end{aligned}}}%
\label{eq:17a} % \nonumber % &
\end{align}}%
%+++++++++++++++++++++++++++++++++++++++++++++
The logarithmic divergence here reflects the fact that $\Delta _{\text{F}}^{(0,0)}(0)$ is not defined. One regularisation suffices to cancel the divergence: the regularised propagator, 
%=============================================
\protect{\begin{align}{{\begin{aligned} 
\Delta _{\text{F}}^{(0,1)}(t) = \int _{C_{\text{F}}} \frac{d \omega }{2\pi}\, 
\frac{\text{e}^{-i\omega t}\,\Omega_1}{(\Omega_0-\omega)(\Omega_1-\omega ) } ,
\end{aligned}}}%
\label{eq:2b} % \nonumber % &
\end{align}}%
%+++++++++++++++++++++++++++++++++++++++++++++
is continuous at $t=0$, and $\Delta _{\text{F}}^{(0,1)}(0) = 0$. 

The regularised propagator and loop can be expressed as legitimate diagrams in {\em another\/} diagram technique, 
%=============================================
\protect{\begin{align}{{\begin{aligned} 
\raisebox{-0.6ex}{\includegraphics{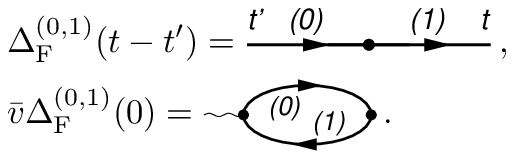}} 
\end{aligned}}}%
\label{eq:23a} % \nonumber % &
\end{align}}%
%+++++++++++++++++++++++++++++++++++++++++++++
This technique comprises four graphical elements, %=============================================
\begin{align} 
\label{eq:39a} % \nonumber % & 
&
\raisebox{-0.6ex}{\includegraphics{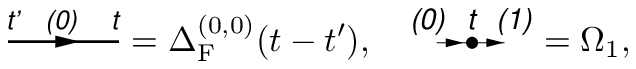}} 
\\ 
\label{eq:40a} % \nonumber % & 
&
\raisebox{-0.6ex}{\includegraphics{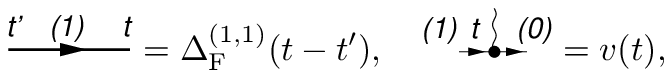}} 
\end{align}%
%+++++++++++++++++++++++++++++++++++++++++++++
where $\Delta _{\text{F}}^{(1,1)}$ is given by Eq.\ (\protect\ref{eq:1b}) with $\Omega_0\to \Omega_1$. Numbers on lines and vertices must match. 
Such diagram technique follows from the time-dependent non-Hermitian Hamiltonian, (see endnote \cite{endn2})
%=============================================
{\begin{multline}\hspace{0.4\columnwidth}\hspace{-0.4\twocolumnwidth} 
\hat {\bar H}_{v}(t) = 
 \Omega_0 \hat b^{\dag} _{(0)} \hat a _{(0)} + 
 \Omega_1 \hat b^{\dag} _{(1)} 
\hat a _{(1)} 
\\ - \Omega_1 \hat b^{\dag} _{(1)} \hat a _{(0)} 
- v(t) \hat b^{\dag} _{(0)} 
\hat a_{(1)} 
, 
\hspace{0.4\columnwidth}\hspace{-0.4\twocolumnwidth} 
\label{eq:13a} % \nonumber % &
\end{multline}}%
%+++++++++++++++++++++++++++++++++++++++++++++
where $\hat a _{(0)},\hat a _{(1)}$ and 
$\hat b^{\dag} _{(0)},\hat b^{\dag} _{(1)}$ 
are {\em dual operator sets\/}, 
%=============================================
\protect{\begin{align}{{\begin{aligned} 
\protect\protect\ensuremath{\big[
\hat a _{(L)}, \hat b^{\dag} _{(K)}
\big]} = \delta _{LK}, \ \ \ L,K =0,1.
\end{aligned}}}%
\label{eq:56a} % \nonumber % &
\end{align}}%
%+++++++++++++++++++++++++++++++++++++++++++++
Heisenberg equations related to (\ref{eq:13a}), 
%=============================================
\protect{\begin{align}{{\begin{aligned} 
\protect\ensuremath{\big(
\Omega _0 - i\partial _t
 \big)} \hat A_{(0)}(t) = v(t) \hat A_{(1)}(t) , \\ 
\protect\ensuremath{\big(
\Omega _1 - i\partial _t
 \big)} \hat A_{(1)}(t) = \Omega_1 \hat A_{(0)}(t) , 
\end{aligned}}}%
\label{eq:29a} % \nonumber % &
\end{align}}%
%+++++++++++++++++++++++++++++++++++++++++++++
implement a natural physical concept: a signal circulating in a feedback loop, with regularisation related to the infinitesimal delay in the loop. 

Diagrams (\ref{eq:39a}), (\ref{eq:40a}) imply that the first two terms in (\ref{eq:13a}) are regarded as free Hamiltonian and the last two as interaction. What if we consider only the last term as interaction and the sum of the first three as free Hamiltonian? To understand what such free Hamiltonian means, we replace the interaction term by the {\em source term\/} $s(t)\hat b ^{\dag}_{(0)}$. Equations (\ref{eq:29a}) then hold with $v(t) \hat A_{(1)}(t)\to s(t)$. Using them we find the {\em linear response function\/} of the system, (cf.\ Fig.\ \ref{CCon})
%=============================================
\protect{\begin{align}{{\begin{aligned} 
\frac{\delta 
\hat A_{(1)}(t)
}{\delta s(t')}\bigg | _{s=0} 
= 
\int _{C_{\text{R}}} \frac{d \omega }{2\pi}\, 
\frac{\text{e}^{-i\omega (t-t')}\,\Omega_1}{(\Omega_0-\omega)(\Omega_1-\omega ) } . 
\end{aligned}}}%
\label{eq:3b} % \nonumber % &
\end{align}}%
%+++++++++++++++++++++++++++++++++++++++++++++
It is instructive to compare this to Kubo's formula \cite{Kubo} 
%=============================================
\protect{\begin{align}{{\begin{aligned} 
\frac{\delta 
\hat A_{(1)}(t)
}{\delta s(t')}\bigg | _{s=0} 
= i \theta(t-t') \protect\protect\ensuremath{\big[
\hat a_{(1)} (t),\hat b ^{\dag}_{(0)} (t')
\big]} 
. 
\end{aligned}}}%
\label{eq:7h} % \nonumber % &
\end{align}}%
%+++++++++++++++++++++++++++++++++++++++++++++
Quantum averaging present in Kubo's expression is omitted here because both sides are c-numbers anyway; the same applies to Eq.\ (\protect\ref{eq:3b}). 
Manipulating complex contours in the conventional way 
\cite{Schw} we recover the two-time commutator, (cf.\ Fig.\ \ref{CCon})
%=============================================
\protect{\begin{align}{{\begin{aligned} 
\protect\protect\ensuremath{\big[
\hat a_{(1)} (t),\hat b ^{\dag}_{(0)} (t')
\big]} 
= 
\int _{C} \frac{d \omega }{2\pi i}\, 
\frac{\text{e}^{-i\omega (t-t')}\,\Omega_1}{(\Omega_0-\omega)(\Omega_1-\omega ) } . 
\end{aligned}}}%
\label{eq:4b} % \nonumber % &
\end{align}}%
%+++++++++++++++++++++++++++++++++++++++++++++
Using the formula, (with $\sigma = 1/\sqrt{1-\Omega_0/\Omega_1}$)
%=============================================
\protect{\begin{align}{{\begin{aligned} 
\frac{\Omega_1}{(\Omega_0-\omega)(\Omega_1-\omega ) } = 
\frac{\sigma ^2}{\Omega_0-\omega} - 
\frac{\sigma ^2}{\Omega_1-\omega} ,
\end{aligned}}}%
\label{eq:5b} % \nonumber % &
\end{align}}%
%+++++++++++++++++++++++++++++++++++++++++++++
we can express the dual operators in terms of the creation and annihilation operators of two harmonic oscillators, 
%=============================================
\protect{\begin{align}{{\begin{aligned} 
&\hat a _{(1)} = \hat a _{\textrm{out}} = \sigma \protect\ensuremath{\big(
\hat a _0 + \hat a _1
 \big)} , \ \ \ 
\hat a _{(0)} = \sigma^{-1} \hat a _0, 
\\
&\hat b^{\dag} _{(0)} = \hat a^{\dag} _{\textrm{in}} = \sigma \protect\ensuremath{\big(
\hat a^{\dag} _0 - \hat a^{\dag} _1
 \big)}, \ \ \ 
\hat b^{\dag} _{(1)} = \sigma^{-1} \hat a^{\dag} _1 . 
\end{aligned}}}%
\label{eq:57a} % \nonumber % &
\end{align}}%
%+++++++++++++++++++++++++++++++++++++++++++++
The alternative notation $\hat a _{(1)} = \hat a _{\textrm{out}}$, $\hat b^{\dag} _{(0)} = \hat a^{\dag} _{\textrm{in}}$ emphasises that these operators correspond to a specially chosen pair of input/output modes of the system. Unlike for a randomly chosen pair, Eq.\ (\protect\ref{eq:3b}) is {\em regular\/}: it lacks the step-function contribution. 
Using (\ref{eq:57a}) it is straightforward to show that the free part in (\ref{eq:13a}) is simply a sum of two oscillator Hamiltonians, 
%=============================================
\protect{\begin{align}{{\begin{aligned} 
\hat {\bar H}_{v}(t)|_{v=0} = 
 \Omega_0 \hat {a}^{\dag} _{0} \hat a _{0} + 
 \Omega_1 \hat {a}^{\dag} _{1} \hat a _{1} .
\end{aligned}}}%
\label{eq:13A} % \nonumber % &
\end{align}}%
%+++++++++++++++++++++++++++++++++++++++++++++

{\em Toy model summary.---\/}%
Regularisation is a result of (a) introducing fictitious degrees of freedom, (b) changing variables in the free Hamiltonian, and (c) modifying the interaction Hamiltonian by replacing a single physical field with a pair of fields with engineered commutational properties. In turn, these are inherently related to response properties of the system. Physically, the emergent regularised system is a {\em quantised low-pass frequency filter with feedback\/}. 

{\em Regularisation algebra.---\/}%
Let $M_0,\cdots,M_N$ ($N\geq 0$) be some real constants (all different). 
Generalising Eq.\ (\protect\ref{eq:5b}) we define a set of complex functions $G_{\textrm{f}}^{(K,L)} (z )$, 
(by default all indices vary from $0$ to $N$) 
%=============================================
\begin{align} 
G_{\textrm{f}}^{(K,L)} ( z ) = % \\ \times 
\left\{
\begin{array}{r}
M_K^{-1} \prod_{l=K}^{L}(1 - z /M _{l})^{-1},\, K\leq L, 
\\ 
\hfill 0, \hfill K>L. 
\end{array}
\right . 
\label{eq:18a} % \nonumber % &
\end{align}%
%+++++++++++++++++++++++++++++++++++++++++++++
and real constants $\varepsilon _K=\pm 1$ and $\sigma _K$ such that 
%=============================================
\begin{align} 
G_{\textrm{f}}^{(0,N)} (z ) = \sum_{K=0}^N \varepsilon _K\sigma _K^2G_{\textrm{f}}^{(K,K)} (z ).
\label{eq:DlR} 
\end{align}%
%+++++++++++++++++++++++++++++++++++++++++++++
For the typical case $M _0\ll M _1\ll\cdots\ll M _N$, 
%=============================================
\protect{\begin{gather}{{\begin{gathered} 
\epsilon_K = (-1)^K, \ K = 0,\cdots,N, \\ 
\sigma _0 \approx \sigma _1 \approx 1, \ \ 
0<|\sigma _K|\ll 1, \ K = 2,\cdots,N. 
\end{gathered}}}
% \nonumber % \eqlabel{} 
\end{gather}}%
%+++++++++++++++++++++++++++++++++++++++++++++
Of use to us will be the recursion relations ($K<L$)
%=============================================
\protect{\begin{align}{{\begin{aligned} 
&\protect\protect\ensuremath{\big[\protect\ensuremath{\big(
M_{L} - z
 \big)}/ M_{L} \big]}G_{\textrm{f}}^{(K,L)}( z ) = G_{\textrm{f}}^{(K,L-1)}( z ), \\
&\protect\protect\ensuremath{\big[\protect\ensuremath{\big(
M_{K} - z
 \big)}/ M_{K+1} \big]}G_{\textrm{f}}^{(K,L)}( z ) = G_{\textrm{f}}^{(K+1,L)}( z ) . 
\end{aligned}}}%
\label{eq:20a} % \nonumber % &
\end{align}}%
%+++++++++++++++++++++++++++++++++++++++++++++

{\em Dirac's filter.---\/}%
Following the pattern of Eqs.\ (\protect\ref{eq:29a}), we wish to replace the equation for the electron operator in spinor quantum electrodynamics by the {\em interacting Dirac low-pass filter\/}, (with $K = 1,\cdots,N$)
%=============================================
\protect{\begin{gather}{{\begin{gathered} 
\protect\ensuremath{\big(
M _0 - i\mbox{$\textstyle\hspace{0.12\pwd} / \hspace{-1\pwd}\hspace{-0.12\pwd}\partial$} 
 \big)} \hat \Psi_{(0)}(x) = e{\hspace{0.4\pwd} / \hspace{-1\pwd}\hspace{-0.4\pwd}\hat A}(x)\hat {\Psi}_{(N)}(x), 
\\ 
\protect\ensuremath{\big(
M _k - i\mbox{$\textstyle\hspace{0.12\pwd} / \hspace{-1\pwd}\hspace{-0.12\pwd}\partial$}
 \big)} \hat \Psi_{(K)}(x) = M _K\hat \Psi_{(K-1)}(x). 
\end{gathered}}}
% \nonumber 
\label{eq:IntF} 
\end{gather}}%
%+++++++++++++++++++++++++++++++++++++++++++++
Here, $\hat \Psi_{(K)}(x)$ are linear combinations of conventional spinor field operators $\hat \Psi_{l}(x)$ with masses $M_l$. 
Feynman's slash notation is used for a scalar product of a 4-vector with Dirac's matrices \cite{Schw}, 
$\mbox{$\textstyle\hspace{0.12\pwd} / \hspace{-1\pwd}\hspace{-0.12\pwd}\partial$} = \gamma ^{\mu }\partial_{\mu } = \gamma ^{\mu }\partial/\partial x^{\mu }$, 
$x = \protect\protect\ensuremath{ \{
t,{\mbox{\rm\boldmath$r$}}
 \}} $, and 
${\hspace{0.4\pwd} / \hspace{-1\pwd}\hspace{-0.4\pwd}\hat A} = \gamma ^{\mu }\hat A_{\mu }$, where $\hat A_{\mu }$ is the 4-vector electromagnetic potential. 
We proceed as for the toy model (\ref{eq:13a}), by firstly changing variables in the free Hamiltonian, and then modifying the interaction. 
The {\em free Dirac low-pass filter\/} is constructed from $N+1$ free Dirac fields $\hat \psi _{l}(x)$ with masses $M_l$: 
%=============================================
\protect{\begin{align}{{\begin{aligned} 
\protect\ensuremath{\big(
M_l - i\mbox{$\textstyle\hspace{0.12\pwd} / \hspace{-1\pwd}\hspace{-0.12\pwd}\partial$}
 \big)} \hat \psi _l(x) = 0, \ l=0,\cdots,N . 
\end{aligned}}}
% \nonumber 
\label{eq:Dir} 
\end{align}}%
%+++++++++++++++++++++++++++++++++++++++++++++
Up to a factor $\gamma ^0$, these follow as Heisenberg equations, 
%=============================================
\protect{\begin{align}{{\begin{aligned} 
i\partial_t\hat \psi _l(x) = \protect\protect\ensuremath{\big[
\hat \psi _l(x) , 
\hat H(\hat {\bar\psi} _l(x),\hat \psi _l(x);M_l)
\big]} . 
\end{aligned}}}%
\label{eq:43a} % \nonumber % &
\end{align}}%
%+++++++++++++++++++++++++++++++++++++++++++++
where $\hat H(\hat {\bar\psi}(x),\hat \psi (x);M)$ is the Hamiltonian of free Dirac's field \cite{Schw}, and $\hat {\bar\psi}_l (x) = \psi^{\dag}_l (x)\gamma ^0$ is Dirac's conjugate of $\hat \psi_l (x)$. 
The dual operator sets comprising the filter are given by recursion relations, ($K=1,\cdots,N$) 
%=============================================
\begin{align} 
&\hat \psi_{(N)} (x)=\hat \psi_{\textrm{out}} (x) = \sum_{l=0} ^N \sigma _l \hat \psi_l (x), 
\label{eq:Def0} 
\\ 
&{\begin{aligned} 
& 
M _K\hat\psi_{(K-1)}(x) = 
\protect\ensuremath{\big(
M _K - i\mbox{$\textstyle\hspace{0.12\pwd} / \hspace{-1\pwd}\hspace{-0.12\pwd}\partial$}
 \big)} \hat\psi_{(K)}(x) 
,
\end{aligned}} \label{eq:50a} % \nonumber % &
\\ 
&\hat {\bar \varphi}_{(0)} (x) = \hat {\bar\psi}_{\textrm{in}} (x) = \sum_{l=0} ^N \epsilon _l\sigma _l \hat {\bar\psi}_l (x) , 
\label{eq:Def1} 
\\ 
&{\begin{aligned} 
& M_{K}\hat{\bar \varphi}_{(K)}(x) = \hat{\bar \varphi}_{(K-1)}(x)\protect\ensuremath{\big(
M_{K-1} + i\loarrow{\mbox{$\textstyle\hspace{0.12\pwd} / \hspace{-1\pwd}\hspace{-0.12\pwd}\partial$}}
 \big)} . 
\end{aligned}}%
\label{eq:44a} % \nonumber % &
\end{align}%
%+++++++++++++++++++++++++++++++++++++++++++++
We now prove the all-important formula for two-time anticommutators, extending Eq.\ (\protect\ref{eq:4b}) to fermions: 
%=============================================
\protect{\begin{align}{{\begin{aligned} 
\protect\protect\ensuremath{\big[
\hat \psi _{(K)}(x),\hat {\bar \varphi }_{(L)} (x') 
\big]}_+ 
= \int _{C} \frac{d^4 p }{i(2\pi)^4}\,\text{e}^{-ip (x-x')} 
G_{\textrm{f}}^{(L,K)} (\mbox{$\textstyle\hspace{0.14\pwd} / \hspace{-1\pwd}\hspace{-0.14\pwd}p$} ) . 
\end{aligned}}}%
\label{eq:21A} % \nonumber % &
\end{align}}%
%+++++++++++++++++++++++++++++++++++++++++++++
where $\mbox{$\textstyle\hspace{0.14\pwd} / \hspace{-1\pwd}\hspace{-0.14\pwd}p$} = \gamma ^{\mu }p_{\mu }$, $p = \protect\protect\ensuremath{ \{
p_0,{\mbox{\rm\boldmath$p$}}
 \}} $. For $K=N,\ L=0$ Eq.\ (\protect\ref{eq:21A}) follows from (\ref{eq:DlR}) and from the complex-contour expression for the unregularised anticommutator \cite{Schw}; for other $K,L$ it is verified by making use of Eqs.\ (\protect\ref{eq:20a}), (\ref{eq:50a}) and (\ref{eq:44a}). For $t=t'$ it reduces to a counterpart of (\ref{eq:56a}), 
%=============================================
\protect{\begin{align}{{\begin{aligned} 
\protect\protect\ensuremath{\big[
\hat \psi _{(K)}(x),\hat {\bar \varphi }_{(L)} (x') 
\big]}_+ |_{t=t'}
= \delta _{KL}\delta ^{(3)}({\mbox{\rm\boldmath$r$}}-{\mbox{\rm\boldmath$r$}}')\gamma ^0 . 
\end{aligned}}}%
\label{eq:45a} % \nonumber % &
\end{align}}%
%+++++++++++++++++++++++++++++++++++++++++++++
The fields $\hat \psi _{(K)}(x)$ hence form a complete set, in the sense that the linear transformation between $\hat \psi _{l}(x)$ and $\hat \psi _{(K)}(x)$ is invertible. Using this we can extend equivalence of the Hamiltonians (\ref{eq:13a}) and (\ref{eq:13A}) to Dirac's fermions:
%\begin{widetext} 
%=============================================
{\begin{multline}\hspace{0.4\columnwidth}\hspace{-0.4\twocolumnwidth} 
\sum_{l=0}^N 
\hat H(\hat {\bar\psi} _l(x),\hat \psi _l(x);M_l)= 
\hat H(\hat {\bar \varphi} _{(0)}(x),\hat \psi _{(0)}(x);M_l) 
\\ + 
\sum_{K=1}^N \protect\protect\ensuremath{\Big[
\hat H(\hat {\bar \varphi} _{(K)}(x),\hat \psi _{(K)}(x);M_l)
\\ 
- M_K 
\hat{\bar \varphi}^{\dag}_{(K)}(x)
\hat\psi_{(K-1)}(x) 
\Big]} 
\hspace{0.4\columnwidth}\hspace{-0.4\twocolumnwidth} 
\label{eq:46a} % \nonumber % &
\end{multline}}%
%+++++++++++++++++++++++++++++++++++++++++++++
%\end{widetext}%
To start with, notice that each successive recursion in (\ref{eq:50a}) shortens the initial sum of operators (\ref{eq:Def0}) by one addend, so that $\hat\psi_{(0)}(x)\propto \hat\psi_{0}(x)$, and 
%=============================================
\begin{gather} 
{\begin{gathered} 
\protect\ensuremath{\big(
M _0 - i\mbox{$\textstyle\hspace{0.12\pwd} / \hspace{-1\pwd}\hspace{-0.12\pwd}\partial$}
 \big)} \hat\psi_{(0)}(x) = 0 . 
\end{gathered}} % \nonumber 
\label{eq:Eq0} 
\end{gather}%
%+++++++++++++++++++++++++++++++++++++++++++++
Equations (\ref{eq:50a}), (\ref{eq:Eq0}) may be used as $N+1$ equations of motion for the filter in place of (\ref{eq:Dir}); assuming otherwise contradicts the fact that $\hat \psi _{(K)}(t)$ form a complete set. 
Furthermore, the RHS of (\ref{eq:46a}) results \cite{endn2} in Eqs.\ (\protect\ref{eq:50a}), (\ref{eq:Eq0}), while the LHS---in Eqs.\ (\protect\ref{eq:Dir}). This is only possible if the RHS and the LHS of (\ref{eq:46a}) coincide. 
The ``free Dirac low-pass filter'' has thus been successfully constructed.

{\em Regularised quantum electrodynamics as quantised filter.---\/}%
%*******************************************
\begin{figure}[t] 
\begin{center} 
\includegraphics[width=0.95\columnwidth]{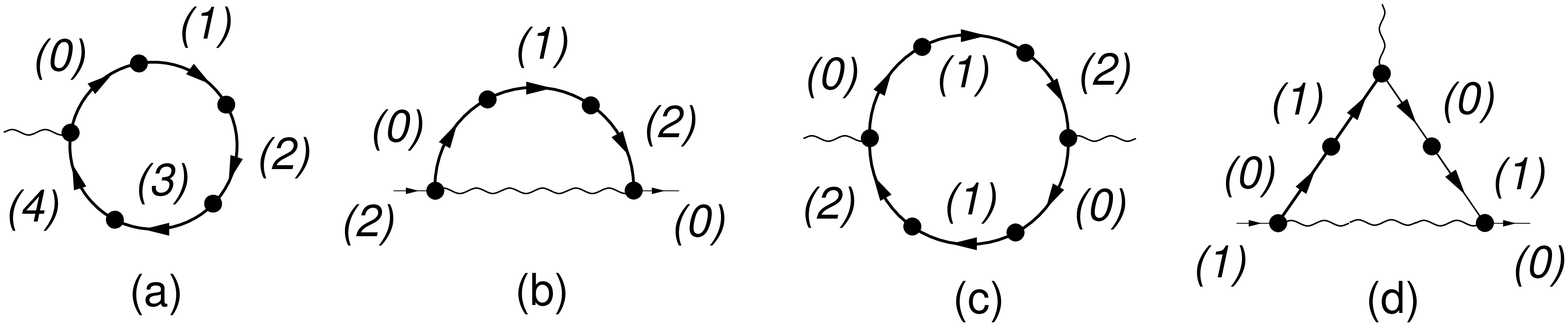}
\end{center} 
\caption{Typical divergent sub-diagrams regularised by the ``Dirac filter.''\label{Conv}} 
\end{figure} 
%*******************************************
To close the ``feedback loop,'' the Dirac filter must be supplemented by the modified electromagnetic interaction. The full Hamiltonian then reads (in IP) 
%=============================================
\begin{align} 
&\hat{\tilde H}_{\textrm{reg}}(t) = 
\hat{\tilde H}_{\textrm{el}}(t) +
\hat{H}_{\textrm{ph}}(t) + \hat{\tilde H}_{\textrm{int}}(t)
\label{eq:63a} % \nonumber % & 
\\
&\hat{\tilde H}_{\textrm{int}}(t) = - e \int d^3{\mbox{\rm\boldmath$r$}}\hat {\bar \psi }_{\textrm{in}} (x) 
\hat{\hspace{0.4\pwd} / \hspace{-1\pwd}\hspace{-0.4\pwd}A}(x)\hat {\psi}_{\textrm{out}} (x) .
\label{eq:Hem} 
\end{align}%
%+++++++++++++++++++++++++++++++++++++++++++++
Here, $\hat{\tilde H}_{\textrm{el}}$ is given by the RHS of (\ref{eq:46a}), and $\hat{H}_{\textrm{ph}}(t)$ is the standard Hamiltonian of the free electromagnetic field \cite{Schw}. Equations (\ref{eq:IntF}) obviously follow from (\ref{eq:Hem}). 
The equation for the electromagnetic field operator may be derived in a conventional way (with non-Hermitian current). 

Typical divergent diagrams regularised by the Dirac filter are shown in Fig.\ \ref{Conv}, where the graphical notation is an obvious generalisation of Eqs.\ (\protect\ref{eq:39a}), (\ref{eq:40a}), cf.\ also endnotes \cite{endn2,endn4} and Eqs.\ (\protect\ref{eq:71a}), (\ref{eq:73a}) below. The electromagnetic propagator \cite{Schw} remains unregularised. It is easy to see that $N=2$ suffices to regularise all physically relevant diagrams, while $N=4$ guaranties convergence of all diagrams, including the vacuum-phase ones and those subject to the Farry theorem \cite{Schw}. Indeed, in any loop the number of electromagnetic propagators cannot exceed the number of electronic ones. Hence the worst possible divergence is exhibited by the ``tadpole'' (Fig.\ \ref{Conv}a); it is regularised with $N=4$. The worst-case scenarios for physical diagrams are the self-mass of the electron (Fig.\ \ref{Conv}b) and the vacuum polarisation (Fig.\ \ref{Conv}c); they become convergent with $N=2$. The vertex correction (Fig.\ \ref{Conv}d) is regular already with $N=1$. 

{\em Pseudo-adjoint and Keldysh techniques.---\/}%
In Eqs.\ (\protect\ref{eq:IntF}), fields $\hat \Psi_{(0)}(x),\cdots,\Psi_{(N-1)}(x)$ may be eliminated resulting in a closed regularised equation for a single field $\hat \Psi(x)=\hat \Psi _{(N)}(x)$: 
%=============================================
{\begin{multline}\hspace{0.4\columnwidth}\hspace{-0.4\twocolumnwidth} 
(M _0 - i\mbox{$\textstyle\hspace{0.12\pwd} / \hspace{-1\pwd}\hspace{-0.12\pwd}\partial$}) (1 - i\mbox{$\textstyle\hspace{0.12\pwd} / \hspace{-1\pwd}\hspace{-0.12\pwd}\partial$}/M _1)\cdots (1 - i\mbox{$\textstyle\hspace{0.12\pwd} / \hspace{-1\pwd}\hspace{-0.12\pwd}\partial$} /M _N)\Psi(t) 
\\ = 
e\hat{\hspace{0.4\pwd} / \hspace{-1\pwd}\hspace{-0.4\pwd}A}(x)\hat {\Psi}(x). 
\hspace{0.4\columnwidth}\hspace{-0.4\twocolumnwidth} 
% \nonumber 
\label{eq:47a} 
\end{multline}}%
%+++++++++++++++++++++++++++++++++++++++++++++
This corresponds to eliminating intermediate vertices (\ref{eq:39a}) in all propagators. 
The RHS in (\ref{eq:47a}) is the usual self-action source. Its action is infinitesimally delayed by the low-pass filter thus cancelling self-action of fermions at infinitesimally small times and distances. 

Intermediate fields can also be eliminated from perturbation calculations. It is convenient to introduce pseudo-conjugation of operators denoted by $^{\ddag}$, and the associated indefinite scalar product denoted as $\langle\cdot ||
\cdot\rangle$, as 
%=============================================
\protect{\begin{align}{{\begin{aligned} 
&{\hat X} ^{\ddag} = \protect{\hat {\mathcal I}}{\hat X}^{\dag}\protect{\hat {\mathcal I}}, \ \ 
\langle\Phi_1 ||\Phi _2\rangle = 
\langle\Phi_1 |\protect{\hat {\mathcal I}}|\Phi _2\rangle , \end{aligned}}}%
\label{eq:33a} % \nonumber % &
\end{align}}%
%+++++++++++++++++++++++++++++++++++++++++++++
where $\left|\Phi _1\right\rangle$, $\left|\Phi _2\right\rangle$ and ${\hat X}$ are arbitrary state vectors and operator in Fock space, $\protect{\hat {\mathcal I}} = 
\epsilon _0^{\hat n_0}\epsilon _1^{\hat n_1}
\cdots\epsilon _N^{\hat n_N}$ is the {\em sign operator\/}, and $\hat n_l$ are number operators for particles with mass $M_l$. All usual properties hold: $
({\hat X}^{\ddag})^{\ddag} = {\hat X},\ 
(a{\hat X}_1+b{\hat X}_2)^{\ddag} = a^*{\hat X}_1^{\ddag}+b^*{\hat X}_2^{\ddag},\ 
({\hat X}_1{\hat X}_2)^{\ddag} = {\hat X}_2^{\ddag}\,{\hat X}_1^{\ddag} 
$, $\langle\Phi_1 ||{\hat X}|\Phi _2\rangle^* = 
\langle\Phi_2 ||{\hat X}^{\ddag}|\Phi _1\rangle $. A diagonal matrix element 
$\langle\Phi ||{\hat X}^{\ddag}{\hat X}|\Phi \rangle^* = 
\langle\Phi |{\hat X}^{\dag}\protect{\hat {\mathcal I}}{\hat X}|\Phi \rangle $ is real but not bound to be positive. For brevity, we shall talk about ${\mathcal I}$-conjugation etc. The sign operator $\protect{\hat{\mathcal I}}$ does not commute with the interaction, so its use is confined to IP. 

By construction, $\hat{\bar\psi }_{\textrm{in}}(x) $ is Dirac's ${\mathcal I}$-conjugate (denoted by tilde) of $\hat{\psi }_{\textrm{out}}(x) \equiv \hat \psi (x)$,
%=============================================
\protect{\begin{gather}{{\begin{gathered} 
\hat{\bar\psi }_{\textrm{in}} (x)
= \sum _{l=0}^N \sigma _l 
\hat {\psi} _l^{\ddag}(x)\gamma ^0 
= \sum _{l=0}^N \sigma _l 
\hat {\tilde\psi} _l(x) 
= \hat{\tilde\psi }(x) . 
\end{gathered}}}%
\label{eq:70a} % \nonumber % &
\end{gather}}%
%+++++++++++++++++++++++++++++++++++++++++++++
The interaction Hamiltonian in IP is ${\mathcal I}$-Hermitian, and the S-matrix in the IP is ${\mathcal I}$-unitary, (cf.\ also endnote \cite{endn2}) 
%=============================================
\protect{\begin{align}{{\begin{aligned} 
\hat S &= T_+ \exp\protect\protect\ensuremath{\bigg[
ie\int d^4 x \hat {\tilde \psi } (x)\hat{\hspace{0.4\pwd} / \hspace{-1\pwd}\hspace{-0.4\pwd}A}(x)\hat {\psi} (x)
\bigg]} , 
\\ 
\hat {\tilde S} &= T_- \exp\protect\protect\ensuremath{\bigg[
-ie\int d^4 x \hat {\tilde \psi } (x)\hat{\hspace{0.4\pwd} / \hspace{-1\pwd}\hspace{-0.4\pwd}A}(x)\hat {\psi} (x)
\bigg]} = \hat S^{-1} 
 . 
\end{aligned}}}%
\label{eq:71a} % \nonumber % &
\end{align}}%
%+++++++++++++++++++++++++++++++++++++++++++++
Wick's theorem ``does not care'' whether $\hat {\psi } (x),\hat {\tilde \psi } (x)$ are genuine conjugates or ${\mathcal I}$-conjugates, hence conventional perturbation calculations apply with the {\em regularised contraction\/}, 
%=============================================
{\begin{multline}\hspace{0.4\columnwidth}\hspace{-0.4\twocolumnwidth} 
-i\protect\protect\ensuremath{\big\langle 0\big|
T_+\hat {\psi } (x)\hat {\tilde \psi } (x')
\big|0\big\rangle} \\ = \int _{C_{\text{F}}} \frac{d^4 p }{(2\pi)^4}\,\text{e}^{-ip (x-x')} 
G_{\textrm{f}}^{(0,N)} (\mbox{$\textstyle\hspace{0.14\pwd} / \hspace{-1\pwd}\hspace{-0.14\pwd}p$} ) , 
\hspace{0.4\columnwidth}\hspace{-0.4\twocolumnwidth} 
\label{eq:73a} % \nonumber % &
\end{multline}}%
%+++++++++++++++++++++++++++++++++++++++++++++
leading directly to regularised diagram series, cf.\ Fig.\ \ref{Conv}. 

The two additional contractions characteristic of Perel-Keldysh's diagram approach \cite{PerKeld}, 
%=============================================
\protect{\begin{align}{{\begin{aligned} 
-i\protect\protect\ensuremath{\big\langle 0\big|
\hat {\psi } (x)\hat {\tilde \psi } (x')
\big|0\big\rangle} , \ \ 
i\protect\protect\ensuremath{\big\langle 0\big|
\hat {\tilde \psi } (x')\hat {\psi } (x)
\big|0\big\rangle} , 
\end{aligned}}}%
\label{eq:73A} % \nonumber % &
\end{align}}%
%+++++++++++++++++++++++++++++++++++++++++++++
are also given by regularised expressions, namely, by Eq.\ (\protect\ref{eq:73a}) with $C_{\text{F}}\to C^{(\pm)}$ (Fig.\ \ref{CCon}). A nontrivial problem is however posed by quantum averaging. 
For example, using conventional perturbation techniques \cite{Schw,PerKeld} we obtain, 
%=============================================
{\begin{multline}\hspace{0.4\columnwidth}\hspace{-0.4\twocolumnwidth} 
\mathcal{G} (x,x';x'',x''') \\ = \text{Tr}\hat\rho\, 
T_- \hat \Psi_{\textrm{out}} (x) \hat {\bar\Psi}_{\textrm{in}} ^{\dag} (x') 
\, T_+ \hat \Psi_{\textrm{out}} (x'') 
\hat {\bar\Psi}_{\textrm{in}} ^{\dag} (x''') 
\\ = \text{Tr}\hat\rho\, 
T_-\hat{\tilde S}\hat \psi (x) \hat {\tilde\psi}(x')
\, T_+ \hat{S}\hat \psi (x'') \hat {\tilde\psi}(x''') . 
\hspace{0.4\columnwidth}\hspace{-0.4\twocolumnwidth} 
\label{eq:36a} % \nonumber % &
\end{multline}}%
%+++++++++++++++++++++++++++++++++++++++++++++
Here, $\hat\rho $ is the Heisenberg density matrix, or, which is the same, the initial state of the system. 
However, a consistent generalisation of the conventional perturbative expression implies also a repalacement of the quantum averaging by ${\mathcal I}$-averaging, 
$
\langle\!\langle \hat X 
\rangle\!\rangle = \text{Tr} \hat\rho\protect{\hat{\mathcal I}}\hat X 
$, resulting in 
%=============================================
{\begin{multline}\hspace{0.4\columnwidth}\hspace{-0.4\twocolumnwidth} 
\mathcal{G}(x,x';x'',x''') \\ = \big\langle\!\big\langle 
T_-\hat{\tilde S}\hat \psi (x) \hat {\tilde\psi}(x')
\, T_+ \hat{S}\hat \psi (x'') \hat {\tilde\psi}(x''') 
\big\rangle\!\big\rangle. 
\hspace{0.4\columnwidth}\hspace{-0.4\twocolumnwidth} 
\label{eq:36A} % \nonumber % &
\end{multline}}%
%+++++++++++++++++++++++++++++++++++++++++++++
This equations follows from (\ref{eq:36a}) if we assume that the initial state contains only physical particles of mass $M_0$, so that $\hat\rho = \hat\rho\protect{\hat{\mathcal I}}$. In turn, this assumption is consistent with the standard one \cite{Schw,PerKeld}, namely, that the interaction is switched on adiabatically, and that in the remote past the system is free. 

It should not be overlooked that we have just postulated an explicit distinction between the past and the future. This raises a question as to what extent regularisations are consistent with reversibility. 

In conclusion, a divergence-free computational scheme has been constructed in the framework of relativistic quantum electrodynamics, and its equivalence to the conventional Feynman-Dyson renormalisation techniques has been demonstrated. 

The author is grateful to R.\ Glauber for triggering his interest in the topic, and to M.\ Efremov and Yu.\ Lozovik for enlightening discussions. 
%******************************************* 
 
%*******************************************

%*******************************************

\begin{thebibliography}{}
\bibitem{endn1}
Strictly speaking, this diagram is excluded by Wick's theorem, so that the whole example is purely illustrative.
\bibitem{endn2}
Quantum dynamics with a time-dependent non-Hermitian Hamiltonian in the Schr\"odinger picture $\hat{H}(t)$ is introduced postulating the evolution operator to be 
\mbox{$\protect{\hat {\mathcal U}}(t)=T_+\exp\big[-i\int_{-\infty}^t\hat H(t')dt'\big]$}. Heisenberg
representation $\protect{\hat {\mathcal X}}(t)=\protect{\hat {\mathcal U}}^{-1}(t)\hat X(t)\protect{\hat {\mathcal U}}(t)$
of a Schr\"odinger operator $\hat X(t)$ obeys the Heisenberg equation $ 
i\partial _t \protect{\hat {\mathcal X}} (t) = 
[
\protect{\hat {\mathcal X}} (t),\protect{\hat {\mathcal H}}(t)
] 
$, where $\protect{\hat {\mathcal H}}(t)=\protect{\hat {\mathcal U}}^{-1}(t)\hat H(t)\protect{\hat {\mathcal U}}(t)$. Such evolution has all the necessary properties of the unitary evolution: it preserves products, 
$\protect{\hat {\mathcal U}}^{-1}(t)\hat{X}_1(t) \hat{X}_2(t) \protect{\hat {\mathcal U}}(t) = 
\protect{\hat {\mathcal X}}_1 (t)
\protect{\hat {\mathcal X}}_2 (t) 
$, and hence all commutators and anticommutators, 
$\protect{\hat {\mathcal U}}^{-1}(t)[\hat{X}_1(t),\hat{X}_2(t)]_{\pm} \protect{\hat {\mathcal U}}(t) = 
[\protect{\hat {\mathcal X}}_1 (t),
\protect{\hat {\mathcal X}}_2 (t)]_{\pm} 
$, allowing one to apply the standard nonstationary perturbation theory to non-Hermitian Hamiltonians.%
\bibitem{endn4}
Phase coefficients for the diagrams are of no importance for regularisation (unlike renormalisation).
\end{thebibliography}
\end{document}